\documentclass[12pt]{iopart}

\usepackage{iopams}
\usepackage{graphicx}

\begin{document}

\title{Electromagnetic force density in dissipative isotropic media}

\author{A Shevchenko, M Kaivola}
\address{Aalto University,
Department of Applied Physics, P.O.~Box 13500, FI-00076 AALTO,
Finland}

\ead{andriy.shevchenko@tkk.fi}

\begin{abstract}
We derive an expression for the macroscopic force density that a narrow-band electromagnetic field imposes on a dissipative isotropic medium. The result is obtained by averaging the microscopic form for Lorentz force density. The derived expression allows us to calculate realistic electromagnetic forces in a wide range of materials that are described by complex-valued electric permittivity and magnetic permeability. The three-dimensional energy-momentum tensor in our expression reduces for lossless media to the so-called Helmholtz tensor that has not been contradicted in any experiment so far. The momentum density of the field does not coincide with any well-known expression, but for non-magnetic materials it matches the Abraham expression.
\end{abstract}

\pacs{42.25.Bs, 03.50.De, 75.80.+q, 77.65.-j}

\maketitle After the first well-known theoretical model for static electromagnetic forces in ponderable media, introduced by H.~von~Helmholtz~\cite{Helmholtz1881}, several other descriptions have been proposed to account for the action of both static and oscillating electromagnetic fields on a medium. The two most famous of them were proposed by H.~Minkowski~\cite{Minkowski1908} and M.~Abraham~\cite{Abraham1909}. Their models give different predictions for the value of the force density in certain particular cases, which has created a century-long scientific debate on the topic~\cite{deGroot-Suttorp1972}--\cite{Barnett2010}. It is well known that both these models ignore the existing electro- and magnetostrictive forces~\cite{Brevik1979,Hakim-Higham1962}. A model by Einstein and Laub~\cite{Einstein-Laub1908} was intended to include these forces, but it turned out to be in disagreement with experiments (see, e.g., Ref.~\cite{Brevik1979} and references therein). A reliable way to also include electro- and magnetostrictive forces can be based on the so-called Helmholtz energy-momentum tensor. Previously we have shown that the Helmholtz tensor can be derived in a relatively simple way starting with the microscopic field-matter interaction picture~\cite{Shevchenko2010}. However, this tensor and the other models discussed so far are not applicable to dissipative media and, therefore, they do not allow calculating electromagnetic forces in many realistic materials.

The aim of this work is to derive a general expression for the macroscopic force density imposed on a dissipative and electrically conducting inhomogeneous isotropic medium by a narrow-band electromagnetic field. We do this by spatially averaging the microscopic Lorentz force density. The result is written in terms of the three-dimensional electromagnetic energy-momentum tensor and the momentum density of the field, which explicitly depend on the complex relative permittivity and permeability of the material. In the limit of lossless medium, the obtained energy-momentum tensor becomes the Helmholtz tensor. We note that the force density expressed in terms of complex-valued quantities has been introduced previously (see, e.g., \cite{Haus1975} and \cite{Kemp2006}), but not in connection with the Helmholtz tensor. The expression obtained by us for the field momentum density does not coincide with any of the well-known expressions, but for lossless dielectrics it converges to an expression obtained by us previously~\cite{Shevchenko2010}, and if the material is not magnetic, it matches the Abraham expression.

The derivation of the force density in dissipative media is similar to that introduced by us in Ref.~\cite{Shevchenko2010} for lossless materials. Since essential new details appear along the derivation, we present it here in its entirety. Within classical electrodynamics, the interaction of an electromagnetic field with matter is unambiguously described by the microscopic Maxwell's equations,
\begin{eqnarray}
&&\epsilon_0\nabla\cdot\textbf{e}=\xi\,,\label{m1}\\
&&\nabla\cdot\textbf{b}=0\,,\label{m2}\\
&&-\nabla\times\textbf{e}=\dot{\textbf{b}}\,,\label{m3}\\
&&\frac{1}{\mu_0}\nabla\times\textbf{b}=\epsilon_0\dot{\textbf{e}}+\textbf{j}\,,\label{m4}
\end{eqnarray}
and the microscopic Lorentz force density
\begin{equation}\label{fl0}
\textbf{f}_{\textrm{mic}}=\xi\textbf{e}+\textbf{j}\times\textbf{b}.
\end{equation}
Here \textbf{e} and \textbf{b} are the microscopic electric and magnetic fields in the medium and $\dot{\textbf{e}}$ and $\dot{\textbf{b}}$ are their time derivatives. The electric charge and current densities $\xi$ and $\textbf{j}$, respectively, can be written in terms of point electric charges $q_i$ as
\begin{eqnarray}
&&\xi=\sum_{i}q_i\delta(\textbf{r}-\textbf{r}_i)\,,\label{xi0}\\
&&\textbf{j}=\sum_{i}q_i\dot{\textbf{r}}_i\delta(\textbf{r}-\textbf{r}_i)\,,\label{j0}
\end{eqnarray}
where $\delta(\textbf{r}-\textbf{r}_i)$ is the Dirac delta function centered at the coordinate $\textbf{r}_i$ of charge $q_i$.

The \emph{bound} electric charges in a medium can be combined into localized groups that belong to individual atoms (or molecules). For each such group one can expand the charge and current densities into Taylor series around the group's center $\textbf{r}_l$ and then truncate the series to include contributions from electric and magnetic dipole moments $\textbf{d}_l^{(\textrm{b})}$ and $\textbf{m}_l^{(\textrm{b})}$ only~\cite{Russakoff1970}. Within this approximation the atoms are treated as point dipoles and the bound charge and current densities in the medium are
\begin{eqnarray}
&&\xi_\textrm{b}=-\sum_{l}\textbf{d}_l^{(\textrm{b})}\cdot\nabla\delta(\textbf{r}-\textbf{r}_l)\,,\label{xi-b0}\\ &&\textbf{j}_\textrm{b}=\sum_{l}\dot{\textbf{d}}_l^{(\textrm{b})}\delta(\textbf{r}-\textbf{r}_l) +\sum_{l}\nabla\times\textbf{m}_l^{(\textrm{b})}\delta(\textbf{r}-\textbf{r}_l)\,,\label{j-b0}
\end{eqnarray}
The moments $\textbf{d}_l^{(\textrm{b})}$ and $\textbf{m}_l^{(\textrm{b})}$ contain the contributions of all bound electric charges of atom $l$. The second term in the expression for $\textbf{j}_\textrm{b}$ originates from the electric current loops due to rotational motion of the charges in the atoms~\cite{Russakoff1970}.

If the medium contains \emph{conduction} charges, an oscillating external field will put these charges into oscillating motion. For an electrically neutral medium, the oscillation of the conduction electrons is accompanied by an out-of-phase oscillation of the residual positive charges of the atoms, which can as well be treated as conduction charges. The oscillation of such positive and negative conduction charges produces their own dipole moments $\textbf{d}_l^{(\textrm{c})}$ and $\textbf{m}_l^{(\textrm{c})}$. If the oscillating electron does not move far from a certain positively charged atom $l$ (which is the case for high-frequency fields), the conduction charge and current densities can be written in the same form as $\xi_\textrm{b}$ and $\textbf{j}_\textrm{b}$, i.e., $\xi_\textrm{c}=-\sum_{l}\textbf{d}_l^{(\textrm{c})}\cdot\nabla\delta(\textbf{r}-\textbf{r}_l)$ and $\textbf{j}_\textrm{c}=\sum_{l}\dot{\textbf{d}}_l^{(\textrm{c})}\delta(\textbf{r}-\textbf{r}_l) +\sum_{l}\nabla\times\textbf{m}_l^{(\textrm{c})}\delta(\textbf{r}-\textbf{r}_l)$.

In order to take into account possible dissipation of electromagnetic energy in the medium, the moments $\textbf{d}_l^{(\textrm{b})}$, $\textbf{m}_l^{(\textrm{b})}$, $\textbf{d}_l^{(\textrm{c})}$, and $\textbf{m}_l^{(\textrm{c})}$ are considered to be \emph{complex vectors}. For example, when calculating an expression for the frequency-dependent dielectric constant $\epsilon$, the contribution of each bound charge $k$ to an atomic dipole moment turns out to be proportional to $e^{-i\omega t}/(\omega_k^2-\omega^2-i\omega\gamma_k)$, while the dipole moment originating from each conduction charge is proportional to $e^{-i\omega t}/(-\omega^2-i\omega\gamma_j)$~\cite{Jackson1975}. Here $\omega_k$ and $\gamma_k$ are the resonance frequency and damping constant of the electron, respectively. For conduction electrons, $\omega_j$ are equal to zero but $\gamma_j$ are not because of ohmic losses. From this point of view the division of electric charges into bound and conduction charges is conditional. The total charge and current densities, $\xi=\xi_\textrm{b}+\xi_\textrm{c}$ and $\textbf{j}=\textbf{j}_\textrm{b}+\textbf{j}_\textrm{c}$, are now given by the complex quantities
\begin{eqnarray}
&&\xi=-\sum_{l}\textbf{d}_l\cdot\nabla\delta(\textbf{r}-\textbf{r}_l)\,,\label{xi-bc}\\
&&\textbf{j}=\sum_{l}\dot{\textbf{d}}_l\delta(\textbf{r}-\textbf{r}_l)+\sum_{l}\nabla\times\textbf{m}_l\delta(\textbf{r}-\textbf{r}_l),\label{j-bc}
\end{eqnarray}
where $\textbf{d}_l=\textbf{d}_l^{(\textrm{b})}+\textbf{d}_l^{(\textrm{c})}$ and $\textbf{m}_l=\textbf{m}_l^{(\textrm{b})}+\textbf{m}_l^{(\textrm{c})}$ are, respectively, the total electric and magnetic dipole moments \emph{per atom}.

When in addition to $\xi$ and $\textbf{j}$ the fields $\textbf{e}$ and $\textbf{b}$ are replaced with complex oscillating functions, the form of the microscopic Maxwell's equations and Eqs.~(\ref{xi-bc}) and (\ref{j-bc}) are preserved, but the Lorentz force density must be rewritten. In terms of complex quantities, one can write the microscopic Lorentz force density averaged over one oscillation period as
\begin{equation}
\bar{\textbf{f}}_{\textrm{mic}}=\frac{1}{2}\textrm{Re}\{\textbf{\textit{f}}_{\textrm{mic}}\}
=\frac{1}{2}\textrm{Re}\{\xi^*\textbf{e}+\textbf{j}^*\times\textbf{b}\},\label{miL-complex}
\end{equation}
where Re denotes the real part, and the asterisk stands for complex conjugation. Here the functions \textbf{e} and \textbf{b} are assumed to oscillate harmonically with slowly varying complex amplitudes, in which case $\Delta\omega\ll\omega$, with $\Delta\omega$ and $\omega$ denoting the field's bandwidth and carrier frequency, respectively. In this case, the changes in the field amplitudes during a single oscillation period of the field are negligibly small. In Eq.~(\ref{miL-complex}) the fast oscillation at the carrier frequency is averaged out and $\bar{\textbf{f}}_{\textrm{mic}}$ remains a slowly varying function of time.

Using Eqs.~(\ref{xi-bc}) and (\ref{j-bc}), the function $\textbf{\textit{f}}_{\textrm{mic}}$ is written as
\begin{eqnarray}
\textbf{\textit{f}}_{\textrm{mic}}\!&=&\!\sum_{l}\Big(\!\!-\textbf{d}_l^*\cdot\nabla\delta(\textbf{r}-\textbf{r}_l)\textbf{e}(\textbf{r})+
\dot{\textbf{d}}_l^*\delta(\textbf{r}-\textbf{r}_l)\times\textbf{b}(\textbf{r})\\\nonumber
&+&(\nabla\times\textbf{m}_l^*\delta(\textbf{r}-\textbf{r}_l))\times\textbf{b}(\textbf{r})\Big).\label{Phi}
\end{eqnarray}
By integrating this function over a representative elementary volume $\delta V$ and dividing the result by this volume, one can find the \emph{macroscopic} force density function $\textbf{\textit{f}}\equiv\langle\textbf{\textit{f}}_{\textrm{mic}}\rangle$, where the angle brackets denote spatial averaging. The function $\textbf{\textit{f}}$ is given by~\cite{Shevchenko2010}
\begin{equation}\label{f1+f2+f3}
\textbf{\textit{f}}=\frac{1}{\delta V}\sum_{l\textrm{ in }\delta V}\left(\nabla(\textbf{d}_l^*\cdot\textbf{e}(\textbf{r}_l))+\nabla(\textbf{m}_l^*\cdot\textbf{b}(\textbf{r}_l))
+\frac{\textrm{d}}{\textrm{d}t}(\textbf{d}_l^*\times\textbf{b}(\textbf{r}_l))\right),
\end{equation}
where the summation is performed over atoms belonging to $\delta V$.

Equation~(\ref{miL-complex}) implies that the \emph{macroscopic} force density averaged over one oscillation period of the field is
\begin{equation}
\bar{\textbf{f}}=\langle\bar{\textbf{f}}_{\textrm{mic}}\rangle=\frac{1}{2}\textrm{Re}\{\textbf{\textit{f}}\}.\label{macL}
\end{equation}
Note that the order of temporal and spatial averaging does not affect the result. For charges within $\delta V$ and for an arbitrary coordinate $\textbf{r}$ in $\delta V$, one can write $\textbf{e}(\textbf{r})=\textbf{e}_{\textrm{ext}}(\textbf{r})+\textbf{e}_{\textrm{int}}(\textbf{r})$ and $\textbf{b}(\textbf{r})=\textbf{b}_{\textrm{ext}}(\textbf{r})+\textbf{b}_{\textrm{int}}(\textbf{r})$, where $\textbf{e}_{\textrm{int}}(\textbf{r})$ and $\textbf{b}_{\textrm{int}}(\textbf{r})$ are the fields produced by the charges in $\delta V$ and the fields $\textbf{e}_{\textrm{ext}}(\textbf{r})$ and $\textbf{b}_{\textrm{ext}}(\textbf{r})$ are created by all sources that are external to $\delta V$. The external fields are independent of the charges and their coordinates in $\delta V$, while the "internal" fields are strongly inhomogeneous around the internal charges. Substituting these expansions into Eq.~(\ref{f1+f2+f3}) and using Eq.~(\ref{macL}), one obtains
\begin{eqnarray}\label{fb-ext+own}
\bar{\textbf{f}}&=&\frac{1}{2}\textrm{Re}\Big\{\frac{1}{\delta V}\sum_{l\textrm{ in }\delta V}\Big(\nabla(\textbf{d}_l^*\cdot\textbf{e}_{\textrm{ext}}(\textbf{r}_l))+\nabla(\textbf{m}_l^*\cdot\textbf{b}_{\textrm{ext}}(\textbf{r}_l))\\\nonumber
&+&\frac{\textrm{d}}{\textrm{d}t}(\textbf{d}_l^*\times\textbf{b}_{\textrm{ext}}(\textbf{r}_l))\Big)\Big\}+\bar{\textbf{f}}_\textrm{int},
\end{eqnarray}
where the second term, $\bar{\textbf{f}}_\textrm{int}$, has the same form as the first one, but with the external fields replaced with the internal ones~\cite{Shevchenko2010}. The term $\bar{\textbf{f}}_\textrm{int}$ is equal to $1/\delta V$ times the total force imposed on all electric charges in $\delta V$ by the spiking fields $\textbf{e}_{\textrm{int}}(\textbf{r})$ and $\textbf{b}_{\textrm{int}}(\textbf{r})$ produced by the charges themselves. According to Newton's third law, this force is equal to zero and thus $\bar{\textbf{f}}_\textrm{int}=0$. Removing $\bar{\textbf{f}}_\textrm{int}$ from Eq.~(\ref{fb-ext+own}) and assuming that in the small volume $\delta V$ the total electric and magnetic dipole moments of each atom are equal to constant vectors \textbf{d} and \textbf{m}, one obtains an expression for $\bar{\textbf{f}}=\textrm{Re}\{\textbf{\textit{f}}\}/2$, where $\textbf{\textit{f}}$ is given by
\begin{eqnarray}\label{f(b)1}
\textbf{\textit{f}}&=&\frac{1}{\delta V}\!\!\sum_{k=x,y,z}\!\!\!\Big(d_k^*\nabla\sum_{l\textrm{ in }\delta V}e_{\textrm{ext},k}(\textbf{r}_l)+m_k^*\nabla\sum_{l\textrm{ in }\delta V}b_{\textrm{ext},k}(\textbf{r}_l)\Big)\\\nonumber
&+&\frac{\textrm{d}}{\textrm{d}t}\Big(\textbf{d}^*\times\frac{1}{\delta V}\sum_{l\textrm{ in }\delta V}\textbf{b}_{\textrm{ext}}(\textbf{r}_l)\Big).
\end{eqnarray}
Here the quantities with subindex $k$ are the Cartesian components of the corresponding vector quantities.

The atoms can to the first order be considered to be distributed uniformly within the small $\delta V$. Since in $\delta V$ the fields $\textbf{e}_{\textrm{ext}}(\textbf{r})$ and $\textbf{b}_{\textrm{ext}}(\textbf{r})$ are smooth charge-independent functions, one can replace the averaging of the fields over the coordinates of the atoms in Eq.~(\ref{f(b)1}) with volume averaging, which results in
\begin{equation}\label{f(b)2}
\bar{\textbf{f}}=\frac{1}{2}\textrm{Re}\Big\{\sum_{k=x,y,z}\left(P_k^*\nabla E_{\textrm{ext},k}+M_k^*\nabla B_{\textrm{ext},k}\right)+\frac{\textrm{d}}{\textrm{d}t}\left(\textbf{P}^*\times\textbf{B}_{\textrm{ext}}\right)\Big\},
\end{equation}
where $\textbf{E}_\textrm{ext}\equiv\langle\textbf{e}_{\textrm{ext}}(\textbf{r})\rangle$ and $\textbf{B}_\textrm{ext}\equiv\langle\textbf{b}_{\textrm{ext}}(\textbf{r})\rangle$. The electric polarization \textbf{P} and magnetization \textbf{M}  are given by
\begin{eqnarray}
&&\textbf{P}\equiv N_{\delta V}\textbf{d}/\delta V=\textbf{P}_\textrm{b}+\textbf{P}_\textrm{c}\,,\label{PP}\\
&&\textbf{M}\equiv N_{\delta V}\textbf{m}/\delta V=\textbf{M}_\textrm{b}+\textbf{M}_\textrm{c}\,,\label{MM}
\end{eqnarray}
with $N_{\delta V}$ denoting the number of atoms in $\delta V$. In the above equations, $\textbf{P}_\textrm{b}$ and $\textbf{M}_\textrm{b}$ originate from bound charges and $\textbf{P}_\textrm{c}$ and $\textbf{M}_\textrm{c}$ from conduction charges. We note that Eq.~(\ref{f(b)2}) holds also for anisotropic and nonlinear materials. In what follows, however, it is assumed that the medium is isotropic and linear, so that the polarization components are described by $\textbf{P}_\textrm{b}=\epsilon_0(\epsilon_\textrm{b}-1)\textbf{E}$ and $\textbf{P}_\textrm{c}=i\textbf{E}\sigma/(\epsilon_0\omega)$, where $\epsilon_b$ is the complex dielectric constant due to bound charges and $\sigma$ the complex electric conductivity; $i$ is the imaginary unity. Usually the polarization is written as $\textbf{P}=\epsilon_0(\epsilon-1)\textbf{E}$, where $\epsilon=\epsilon_\textrm{b}+i\sigma/(\epsilon_0\omega)$ is the overall dielectric constant~\cite{Jackson1975}. Similarly, the magnetization can be written as $\textbf{M}=(\mu-1)\textbf{H}$, with $\mu$ being the overall complex relative permeability of the medium. The above equations are valid, if the size $\delta V^{1/3}$ is much larger than the distance $l_\textrm{e}$ over which a conduction electron moves during one oscillation period of the field. This condition limits the theory to high-frequency fields. However, even for copper in an electric field with an amplitude of 1~kV/cm, the distance $l_\textrm{e}$ is on the order of 1~$\mu$m or less already at a frequency of 1~GHz.

The averaged external electric field can be found from $\textbf{E}_\textrm{ext}=\textbf{E}-\langle\textbf{e}_{\textrm{int}}\rangle$~\cite{Shevchenko2010}. For an \emph{arbitrary} charge distribution in $\delta V$, the field $\langle\textbf{e}_{\textrm{int}}\rangle$ is equal to $-\textit{\textbf{D}}_{\delta V}/(3\epsilon_0\delta V)$, where $\textit{\textbf{D}}_{\delta V}=\textbf{P}\delta V$ is the total dipole moment of the medium within $\delta V$ (see also sect.~2.13 in~\cite{Lorrain-Corson1970}). Therefore, the external field is
\begin{equation}\label{E--ext}
\textbf{E}_{\textrm{ext}}=\textbf{E}+\textbf{P}/(3\epsilon_0)=\textbf{E}(\epsilon+2)/3,
\end{equation}
which has the same form as the traditional local field with the Lorentz correction. Note, however, that the polarization $\textbf{P}$ also contains the contribution from the electric dipole moments of the conduction charges. Similarly, the external magnetic field is calculated as $\textbf{B}_\textrm{ext}=\textbf{B}-\langle\textbf{b}_{\textrm{int}}\rangle$. The field $\langle\textbf{b}_{\textrm{int}}\rangle$ created by microscopic electric current loops in $\delta V$ is given by $\langle\textbf{b}_{\textrm{int}}\rangle=2\mu_0\textbf{M}/3$ (see Eq.~(9-22) in \cite{Lorrain-Corson1970}), which yields
\begin{equation}\label{B--ext}
\textbf{B}_{\textrm{ext}}=\textbf{B}-2\mu_0\textbf{M}/3=\mu_0\textbf{H}(\mu+2)/3,
\end{equation}
where the relation $\textbf{B}=\mu_0\mu\textbf{H}$ has been used. Substituting the calculated $\textbf{E}_\textrm{ext}$ and $\textbf{B}_{\textrm{ext}}$ into Eq.~(\ref{f(b)2}) and expressing \textbf{P} and \textbf{M} through \textbf{E} and \textbf{H}, one obtains the force density
\begin{eqnarray}\label{f-f-f}
\bar{\textbf{f}}=\frac{1}{2}&\textrm{Re}\Big\{&\frac{\epsilon_0(\epsilon^*-1)}{3}[|E|^2\nabla\epsilon+(\epsilon+2)\sum_{k=x,y,z}E_k^*\nabla E_k]\\\nonumber
&+&\frac{\mu_0(\mu^*-1)}{3}[|H|^2\nabla\mu+(\mu+2)\sum_{k=x,y,z}H_k^*\nabla H_k]\\\nonumber
&+&\frac{\partial}{\partial t}\left(\frac{\epsilon_0\mu_0(\epsilon^*-1)(\mu+2)}{3}\textbf{E}^*\times\textbf{H}\right)\Big\}
\end{eqnarray}
that depends on \textbf{E} and \textbf{H} and on the complex parameters $\epsilon$ and $\mu$.

The macroscopic fields \textbf{E} and \textbf{H} satisfy the macroscopic Maxwell's equations
\begin{eqnarray}
&&\nabla\cdot\textbf{D}=0\,,\label{M1}\\
&&\nabla\cdot\textbf{B}=0\,,\label{M2}\\
&&-\nabla\times\textbf{E}=\dot{\textbf{B}}\,,\label{M3}\\
&&\nabla\times\textbf{H}=\dot{\textbf{D}}\,,\label{M4}
\end{eqnarray}
where the electric charge and current densities due to conduction charges are absorbed in $\textbf{D}=\epsilon_0\epsilon\textbf{E}$. With the help of these equations and using standard differential identities involving vectors and dyads, the obtained equation for $\bar{\textbf{f}}$ can be rewritten in a general form as
\begin{equation}\label{f}
\bar{\textbf{f}}=-\nabla\cdot\hat{\overline{\textbf{T}}}-\frac{\textrm{d}\bar{\textbf{G}}}{\textrm{d}t},
\end{equation}
where the energy-momentum tensor $\hat{\overline{\textbf{T}}}$ and the momentum density $\bar{\textbf{G}}$ are given by
\begin{eqnarray}
\hat{\overline{\textbf{T}}}&=&\frac{\epsilon_0}{2}(-\textrm{Re}\{\epsilon^*\textbf{E}^*\textbf{E}\}+
\frac{2+2\textrm{Re}\{\epsilon\}-|\epsilon|^2}{6}|E|^2\hat{\textbf{I}})\label{T}\\\nonumber
&+&\frac{\mu_0}{2}(-\textrm{Re}\{\mu^*\textbf{H}^*\textbf{H}\}+\frac{2+2\textrm{Re}\{\mu\}-|\mu|^2}{6}|H|^2\hat{\textbf{I}})\,,\\
\bar{\textbf{G}}&=&\frac{1}{6c^2}\textrm{Re}\{(2+2\epsilon^*\mu-2\epsilon^*+\mu)\textbf{E}^*\times\textbf{H}\}.\label{G}
\end{eqnarray}
Here, $\textbf{E}^*\textbf{E}$ and $\textbf{H}^*\textbf{H}$ denote the outer products of the vectors, $E$ and $H$ are the complex amplitudes of the fields, $\hat{\textbf{I}}$ denotes the unit tensor, and $c$ is the speed of light in vacuum. The form of Eq.~(\ref{f}) is not only physically insightful, but also very convenient in view of calculations, since for stationary fields the force on an object in a medium can be calculated simply by integrating $\textbf{n}\cdot\hat{\overline{\textbf{T}}}$ over the \emph{surface} inclosing the object instead of integrating $\bar{\textbf{f}}$ over the volume of the object; \textbf{n} is the unit vector normal to the surface. It can be seen that if the medium is lossless, so that both $\epsilon$ and $\mu$ are real, Eqs.~(\ref{T}) and (\ref{G}) converge to Eqs.~(35) and (36) in~\cite{Shevchenko2010}. The tensor in Eq.~(35) of Ref.~\cite{Shevchenko2010} is the Helmholtz tensor (see, e.g., Ref.~\cite{Brevik1979}), and the field momentum density of Ref.~\cite{Shevchenko2010} still has to be experimentally verified. At high frequencies, $\mu$ is equal to 1 for most materials, and the momentum density $\bar{\textbf{G}}$ becomes
\begin{equation}\label{G-opt}
\bar{\textbf{G}}_{\mu=1}=\frac{1}{2}\textrm{Re}\Big\{\frac{1}{c^2}\textbf{E}^*\times\textbf{H}\Big\}
\end{equation}
independently of $\epsilon$. Equation~(\ref{G-opt}) is seen to match the Abraham's expression for $\textbf{G}$ averaged over one oscillation period of the field within the slowly varying envelope approximation. Since Eqs.~(\ref{f})-(\ref{G}) are valid for dissipative media, they can be applied also to \emph{metamaterials} that can have $\mu\neq1$ even at optical frequencies. The quantity $\textrm{d}\bar{\textbf{G}}/\textrm{d}t$ is equal to zero for constant-amplitude harmonic fields, and in the case of a slowly varying amplitude the contribution of this quantity to $\bar{\textbf{f}}$ is small compared to the contribution of $\nabla\cdot\hat{\overline{\textbf{T}}}$.

It is worth mentioning that if one deals with dielectric liquids at hydrodynamic equilibrium with the field, the negative of the gradient  of the hydrostatic excess pressure can be found to compensate for the electromagnetic strictive force density at each point in the liquid. Inclusion of this hydrostatic force density often leads to a result that can as well be obtained by using Minkowksi's or Abraham's tensors~\cite{Brevik1979}. However, in a non-equilibrium case, e.g., immediately after introducing the field in the medium, the pressure evolves and \emph{does not} compensate for the electromagnetic strictive force.

As a simple example of new phenomena that can be revealed by applying Eqs.~(\ref{f})-(\ref{G}), let us consider a monochromatic electromagnetic plane wave ($\textrm{d}\bar{\textbf{G}}/\textrm{d}t=0$) that propagates along \textbf{z}-axis in a dissipative medium and has an attenuation length of $z_\textrm{a}$. At $z=0$, the complex amplitudes of the field are $E_0$ and $H_0$. The time-averaged force imposed by the field on the part of the medium that is confined between the planes $z=0$ and $z=z'$ is
\begin{equation}\label{F1}
\textbf{F}=-\int_{z=0}\textbf{n}_1\cdot\hat{\overline{\textbf{T}}}\textrm{d}x\textrm{d}y-\int_{z=z'}\textbf{n}_2\cdot\hat{\overline{\textbf{T}}}\textrm{d}x\textrm{d}y,
\end{equation}
where the surface integrals are obtained by applying Gauss' integration law to the original volume integral; $\textbf{n}_1$ and $\textbf{n}_2$ are the normals to the integration surfaces directed outward from the section of the medium in question. If \textbf{E} and \textbf{H} are everywhere perpendicular to $\textbf{z}$ and $z'\gg z_a$, Eq.~(\ref{F1}) becomes
\begin{eqnarray}\label{F2}
\textbf{F}&=&\hat{\textbf{z}}\Big\{\frac{2+2\textrm{Re}\{\epsilon\}-|\epsilon|^2}{6}\int_{z=0}|E_0|^2\textrm{d}x\textrm{d}y\\\nonumber
&+&\frac{2+2\textrm{Re}\{\mu\}-|\mu|^2}{6}\int_{z=0}|H_0|^2\textrm{d}x\textrm{d}y\Big\},
\end{eqnarray}
where Eq.~(\ref{T}) has been used; $\hat{\textbf{z}}$ is the unit vector along \textbf{z}. The integrals in Eq.~(\ref{F2}) are proportional to the intensity $I_0$ of the plane wave at $z = 0$, since $|E_0|^2=2I_0\sqrt{\mu_0/\epsilon_0}|\mu|/\textrm{Re}\{\sqrt{\epsilon\mu}\}$ and
$|H_0|^2=2I_0\sqrt{\epsilon_0/\mu_0}|\epsilon|/\textrm{Re}\{\sqrt{\epsilon\mu}\}$. In terms of $I_0$, the force \textbf{F} per unit cross-sectional area of the wave, pressure \textbf{p}, reads
\begin{eqnarray}\label{F4}
\textbf{p}&=&\frac{\hat{\textbf{z}}I_0}{c\textrm{Re}\{\sqrt{\epsilon}\}}\frac{2+2\textrm{Re}\{\epsilon\}-|\epsilon|^2+3|\epsilon|}{6},
\end{eqnarray}
where the medium is assumed to be non-magnetic so that $\mu\equiv1$.

It can be seen that for sufficiently large $|\epsilon|$, the pressure \textbf{p} becomes directed \emph{opposite} to the propagation direction of the beam. Thus, if one would consider the matter-field interaction purely in terms of momentum exchange between photons in the wave and the medium, the obtained average momentum per photon would be negative. Obviously, photons can not only share their momenta with the medium, but also impose gradient forces on it. The pressure \textbf{p} in Eq.~(\ref{F4}) contains both a positive component due to the momentum exchange (the usual radiation-pressure component) and a negative component due to the intensity gradient originating from the attenuation of the wave. For large $|\epsilon|$, the second contribution can be stronger than the first one. Single-crystal silicon can be considered as an example. For a wavelength of, say, 532~nm (frequency-doubled Nd:YAG laser), we have $\epsilon\approx20.3 + i 1.0$, $\textrm{Re}\{\sqrt{\epsilon}\}\approx4.5$, and $|\epsilon|\approx20.3$~\cite{HCP}. Substituting these values into Eq.~(\ref{F4}) yields $\textbf{p}=-11.4\hat{\textbf{z}}I_0/c$. This pressure points against $\hat{\textbf{z}}$ and is 11.4 times higher than a pressure that would result from full absorption of the same wave by an object in vacuum. Another example is salt water at 1-THz field frequency. For a mass fraction of 0.25 of NaCl in the solution, the solution is characterized by $\epsilon\approx4.9 + i 3.1$, $\textrm{Re}\{\sqrt{\epsilon}\}\approx2.3$, and $|\epsilon|\approx5.8$~\cite{Jepsen2009}. The resulting pressure is $\textbf{p}=-0.3\hat{\textbf{z}}I_0/c$ that is also opposite to \textbf{z}.

Suppose now that an electromagnetic beam instead of a plane wave is interacting with the previous medium of salt water. If the beam diameter is large compared to $z_a$, then switching on the beam will lead to a motion of the liquid toward the beam source. In time, the hydrostatic pressure will be redistributed to compensate for the gradient force density due to the field. However, the positive radiation-pressure force component will remain uncompensated and, hence, the liquid will eventually flow along the beam axis in the positive direction of $\textbf{z}$. In principle, the same motion of the hydrodynamically equilibrated liquid can be obtained also within Minkowski's and Abraham's theories written in the domain of complex functions, but what cannot be obtained using these theories is the intensity dependent hydrostatic pressure. This is explained by the fact that Minkowski's and Abraham's force densities implicitly contain the action of the medium on itself in the form of an Archimedes-like force density that at equilibrium compensates for the compressive action of the field.

In conclusion, we have derived the equation for the force density imposed by a narrow-band electromagnetic field on a medium that is characterized by complex-valued electric permittivity and magnetic permeability. Equation~(\ref{f}) expresses the force density in the form that allows one to conveniently calculate the overall force imposed by the field on an arbitrary part of the medium by evaluating a surface instead of a volume integral, if the field is stationary. For narrow-band fields, Eq.~(\ref{G}) can be used to evaluate the momentum density of the field averaged over its single oscillation pediod. This quantity depends on $\epsilon$ and $\mu$, but it takes on the Abraham's form if the material is not magnetic, which is the case for essentially all materials in high-frequency fields. However, Eqs.~(\ref{f})-(\ref{G}) can be applied also to high-frequency magnetic metamaterials, which can reveal unexpected phenomena associated with electromagnetic forces.

\ack We acknowledge financial support from the Academy of Finland and thank Prof. B.~J.~Hoenders for useful discussions on the subject.


\section*{References}
\begin{thebibliography}{30}

\bibitem{Helmholtz1881}
von~Helmholtz~H 1881 {\it Wied. Ann.} {\bf 13} 385.

\bibitem{Minkowski1908}
Minkowski H 1908 {\it Nachr. Ges. Wiss. G\"ottingen} 53; 1910 {\it Math. Ann.} {\bf 68} 472.

\bibitem{Abraham1909}
Abraham M 1909 {\it Rend. Circ. Matem. Palermo} {\bf 28} 1.

\bibitem{deGroot-Suttorp1972}
de~Groot S R and Suttorp L G 1972 {\it Foundations of Electrodynamics} (North-Holland Publ. Corp., Amsterdam).

\bibitem{Gordon1973}
Gordon J P 1973 {\it Phys. Rev. A} {\bf 8} 14.

\bibitem{Peierls1976}
Peierls R 1976 {\it Proc. Roy. Soc. London, Ser. A} {\bf 347} 475.

\bibitem{Brevik1979}
Brevik I 1979 {\it Phys. Rep.} {\bf 52} 133.

\bibitem{Lai-Suen-Young1982}
Lai H M, Suen W M and Young K 1982 {\it Phys. Rev.} {\bf 25} 1755.

\bibitem{Nelson1991}
Nelson D F 1991 {\it Phys. Rev. A} {\bf 44} 3985.

\bibitem{Raabe-Welsch2005}
Raabe C and Welsch D -G 2005 {\it Phys. Rev. A} {\bf 71} 013814.

\bibitem{Kemp2007}
Kemp B A, Kong J A and Grzegorczyk~T~M 2007 {\it Phys.~Rev.~A} {\bf 75} 053810.

\bibitem{Mansuripur2007}
Mansuripur M 2007 {\it Opt. Express} {\bf 15} 13502; Mansuripur M 2008 {\it Opt. Express} {\bf 16} 5193.

\bibitem{Pfeifer2007}
Pfeifer R N C, Nieminen T A, Heckenberg N R, Rubinztein-Dunlop H 2007 {\it Rev. Mod. Phys.} {\bf 79} 1197.

\bibitem{Mansuripur2010}
Mansuripur M 2010 {\it Opt. Commun.} {\bf 283} 1997.

\bibitem{Barnett2010}
Barnett S M 2010 {\it Phys. Rev. Lett.} {\bf 104} 070401.

\bibitem{Hakim-Higham1962}
Hakim S S and Higham J B 1962 {\it Proc. Phys. Soc.} {\bf 80} 190.

\bibitem{Einstein-Laub1908}
Einstein A and Laub J 1908 {\it Ann. Physik} {\bf 26} 541.

\bibitem{Shevchenko2010}
Shevchenko A, Hoenders B J 2010 {\it New~J.~Phys.} {\bf 12} 053020.

\bibitem{Haus1975}
Haus H A, Kogelnik H 1975 {\it J. Opt. Soc. Am.} {\bf 66} 320.

\bibitem{Kemp2006}
Kemp B A, Grzegorczyk T M, Kong J A {\it Phys. Rev. Lett.} {\bf 97} 133902.

\bibitem{Russakoff1970}
Russakoff G 1970 {\it Am. J. Phys.} {\bf 38} 1188.

\bibitem{Jackson1975}
Jackson J D 1975 {\it Classical Electrodynamics, sect.~7.5} (John Wiley \& Sons, New York).

\bibitem{Lorrain-Corson1970}
Lorrain P and Corson D 1970 {\it Electromagnetic Fields and Waves} (W.~H.~Freeman and Company, San Francisco).

\bibitem{HCP}
Lide D R 1996 {\it CRC Handbook of Chemistry and Physics} (CRC Press, USA).

\bibitem{Jepsen2009}
Jepsen P U, Merbold H 2010 {\it J Infrared Milli Terahz Waves} {\bf 31} 430.

\end{thebibliography}
\end{document}